\documentclass[aps,prl,twocolumn,showpacs,floatfix,superscriptaddress,reprint]{revtex4-1}
\usepackage[T1]{fontenc}
\usepackage{amsmath,amssymb,amsfonts,bbm,hyperref,times}
\usepackage[dvips,usenames]{color}
\usepackage{graphicx}

\begin{document}
\title{Dynamical recurrence and the quantum control of coupled oscillators}
\author{Marco G. Genoni}
\affiliation{QOLS, Blackett Laboratory, Imperial College London, London SW7 2BW, UK}
\author{Alessio Serafini}
\affiliation{Department of Physics and Astronomy, University College London, Gower Street, London WC1E 6BT, UK}
\author{M. S. Kim}
\affiliation{QOLS, Blackett Laboratory, Imperial College London, London SW7 2BW, UK}
\author{Daniel Burgarth}
\affiliation{Institute of Mathematics and Physics, Aberystwyth University, Aberystwyth  SY23 3BZ, United Kingdom}
\begin{abstract}
Controllability 
-- the possibility of performing any target dynamics by applying a set of available operations -- 
is a fundamental requirement for the practical use of any physical system.
For finite-dimensional systems, such as spin systems, 
precise criteria to establish controllability,  such as the so-called
{\em rank criterion}, are well known. 
However most physical systems require a description in terms of an 
infinite-dimensional Hilbert space whose controllability properties are poorly understood.
Here, we investigate infinite-dimensional bosonic 
quantum systems -- encompassing quantum light, ensembles of bosonic atoms, 
motional degrees of freedom of ions, and nano-mechanical oscillators --
governed by quadratic Hamiltonians (such that their evolution is analogous to 
coupled harmonic oscillators).
After having highlighted the intimate connection between controllability 
and recurrence in the Hilbert space, we prove that, for coupled oscillators, 
a simple extra condition has to be fulfilled to extend the {\em rank
criterion} to infinite-dimensional quadratic systems.  
Further, we present a useful application of our finding, by proving indirect 
controllability of a chain of harmonic oscillators.
\end{abstract}
\maketitle
One of the most fundamental questions  in science is what 
kind of dynamics a given system can host. 
Control theory addresses this question in the light of how 
the dynamics of the system can change as a response to 
our attempts of steering it. 
Control theory can be applied at different levels:
when dealing with computing devices, for example, 
one could either classify their dynamics by the primary logical operations 
they can perform or, at a higher and arguably more useful level, 
by determining what kind of programs they can run.
In quantum computing the first level is typically determined by 
the experiments, and provides one with a description of
the Hamiltonian of the system under consideration. 
On the other hand the second level corresponds to the 
set of quantum algorithms that the quantum
computer is capable of running. 
It is at this second level that the capability
for a device to perform the algorithms 
theorists dream of is established or disproved.
\par
\begin{figure}
\includegraphics[width=0.95\columnwidth]{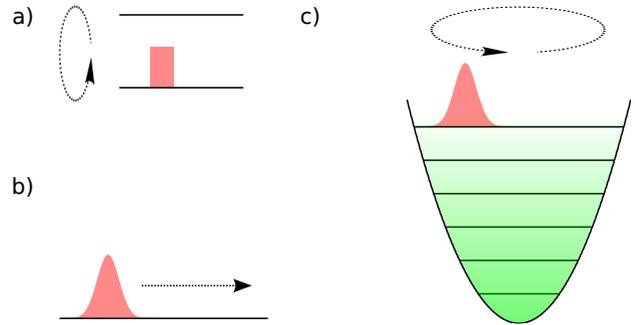}
\caption{
In finite dimensions, quantum systems recur (a), while,
in infinite dimensions, single-directed movement can carry
an `intrinsic clock' and systems do not recur anymore (b). However,
under certain conditions, we prove that harmonic oscillator systems (c) 
can recur as if they were finite-dimensional.
\label{f:recurrence}
}
\end{figure}

To connect the experimental and theoretical 
levels, one faces the problem of translating the Hamiltonian description 
to a description in terms of algorithms it can perform. 
For finite-dimensional quantum systems, for instance {\em qubits} and {\em qudits}, 
this translation has been accomplished in the 1970s by the development 
of an elegant mathematical framework
dubbed as algebraic control \cite{JS,Dalessandro}.
In infinite dimension, however, such a translation has been so far elusive. 
Roughly speaking, the problem encountered is described as 
follows (see Fig. \ref{f:recurrence}): 
in finite-dimensional systems, the state space is `limited', 
and if the quantum state evolves in one specific direction, 
it will eventually return to where it started. 
Classically this would be Poincare's celebrated recurrence theorem, 
whose quantum mechanical counterpart is given in \cite{QRT}. Hence, in finite 
dimension, there is in some sense no need to distinguish between opposite directions 
or, equivalently, to specify the direction of time. For control theory
this allows one to develop a picture in which time is fully eliminated. 
Single-directed movement in infinite-dimensional systems however can carry an `intrinsic clock', 
e.g. distance travelled from some initial state, and time cannot be eliminated.
\par
Now in order to assess quantum control of an infinite-dimensional system,
our question is: do all infinite-dimensional systems have an intrinsic clock? 
A counterexample is given by harmonic oscillators: just by looking 
at the system state of an oscillator, one cannot tell how long it has 
been running from any specific initial state. 
Hence, there is hope to perform the same time elimination 
and reach an easy description of operations that can be performed in 
quantum harmonic oscillators. 
Note that these systems are not mere theoretical curiosities. 
The control of the infinite-dimensional degrees of freedom of light, 
trapped particles, nano- and opto-mechanical oscillators, superconductors, 
Bose-Einstein condensates, and of
collective spins of atomic vapors or solid-state devices are all of 
major technological interest, and the primary way to address most of such 
degrees of freedom is the manipulation of quadratic Hamiltonians,
which correspond to descriptions in terms of quantum harmonic oscillators
\cite{light,ions1,ions2,nano,opto,ensembles}.
We will see a specific example concerning the control of 
arrays of trapped ions at the end of the Letter.
\par
In this letter, we prove that a restrictive condition has to be fulfilled
in order to assess the controllability of quantum harmonic oscillator networks.
We shall observe that such systems share
substantial similarities with finite-dimensional dynamics, which explains 
the success of previous numerical results \cite{rabitz}. We also 
demonstrate the potential impact of our findings by showing how indirect 
control methods \cite{Daniel},
developed previously for spin systems only, can also be applied to 
oscillators, possibly leading to resource efficient cooling and control protocols.
We start by presenting the basic notions of the algebraic control, 
revisiting the proof of the {\em Lie algebra rank criterion}  \cite{JS,Dalessandro}
and finding why it fails to be sufficient for the controllability
of generic infinite-dimensional systems. Then we will focus on quadratic bosonic
Hamiltonians, which give rise to the so-called {\em Gaussian operations}, and
we will determine a condition such that the {\em rank
criterion} will still be sufficient for controllability in the restricted Gaussian sense. 
In the end, we will present an example relevant to arrays of trapped ions and to
chain of nano-mechanical oscillators, where this condition is fulfilled and 
where local controllability can be proven by applying our general analysis.

\par
{\em Algebraic control} - 
Let us start by reviewing a finite-dimensional control setup 
in quantum physics.
Suppose an experimentalist succeeds in setting up a system described by the Hamiltonian
\begin{equation}
H(t) = H_0 + \sum_{k=1}^m f_k(t) H_k, \label{eq:Ham}
\end{equation}
where the $H_{k}$
are a set of controlling Hamiltonians that can be switched on and off.
The Schr\"odinger equation for the time evolution operator $U$ then reads
\begin{align}
\frac{dU}{dt} = -i H(t)\: U\qquad U(0)=\mathbbm{1}. \label{eq:Uni}
\end{align}

The main goal of a control theorist is to determine which quantum algorithms,
{\em i.e.} which unitary operators $U$, the experimentalist can, in principle, achieve by 
setting the right switching times for the $f_k$.\\
In this Letter we are interested in controlling systems described as 
coupled oscillators, thus we need to introduce some additional notation and terminology. 
We shall consider an $n$-mode bosonic system, described by $n$ pair of 
quadrature operators $q_j$ and $p_j$ satisfying the canonical commutation
relation $[q_k,p_l] = i\delta_{k,l}$. By introducing the vector of operators
$R^T = (q_1,p_1,\dots,q_n,p_n)$, the commutation relation can be
written as
$[R_k, R_l] = i \Omega_{kl}$  where $\Omega$ is the $(2n)\times (2n)$
symplectic form whose matrix elements are 
$\Omega_{jk}=\delta_{j+1,k}[1-(-1)^j]/2-\delta_{j,k+1}[1+(-1)^j]/2$
in terms of Kronecker deltas $\delta_{j,k}$.\\
In particular we will consider systems described and controlled by 
Hamiltonians that are bilinear in the quadrature operators, {\em i.e.} that can be written as
$H = (1/2) \sum_{k,l} A_{kl} R_k R_l$
where $A$ are real and symmetric $2n\times 2n$ matrices.
The corresponding evolution operators in the infinite-dimensional
Hilbert space, defined as $U=e^{-i H t}$, are the so-called 
Gaussian unitary operations since they preserve the
Gaussian character of quantum states \cite{Gaussians1,Gaussians2}.
If we consider the Heisenberg evolution for the quadrature operators 
vector $R$, we obtain the equation $U^{\dag} R U = S R$ 
where $S=e^{-A \Omega t}$ is a $2n\times 2n$ matrix belonging to 
the real symplectic group $Sp(2n,\mathbbm{R})$ satisfying the 
equation $S\Omega S^T =\Omega$  \cite{agarwal}. 
Then, restricting to quadratic Hamiltonians, 
there is a one-to-one correspondence between the evolution
operator in the infinite-dimensional Hilbert space $U$ and the 
finite-dimensional matrix $S$, in particular they correspond to different 
representations of the real symplectic group \cite{Arwind}. 
As a consequence  Eq.~(\ref{eq:Uni}) 
can be recast in terms of the {\em symplectic} 
representation, providing one with the following evolution equation
\begin{align}
\frac{dS}{dt} &= G(t) S \qquad 
S(0) = \mathbbm{1} \label{eq:evolution}, 
\end{align}
where, as pointed out above, $S$ is no longer unitary, and
$G(t)=-(A^{(0)} + \sum_{k=1}^m f_k(t) A^{(k)}) \Omega$ is no longer anti-hermitian.
In this framework, the main goal of the control theorist can be summed up 
by the following question:
which symplectic time-evolutions $S$ are achievable by controlling $G(t)$ 
via the functions $f_k(t)$?\\
Because the solutions of Eq.~(\ref{eq:evolution}) are elements of matrix groups, one
can apply the beautiful framework of Lie groups to tackle such questions.
 Let us suppose that, by setting the control functions equal to 
constant values in $G(t)$,  we can identify a set  
$\mathcal{E}=\{  \widetilde{G}_1,\cdots, \widetilde{G}_m\}$ 
of linearly independent generators of a Lie algebra $\mathcal{L}$.
This assumption is known as the {\em Lie algebra rank 
criterion}. Its relevance is due to the fact that,
if the corresponding Lie group $\mathcal{G}=e^{\mathcal{L}}$ is 
a subset of a compact group, then
all its elements can be implemented with arbitrary precision
and the system is said to be controllable.
The {\em Lie algebra rank criterion} is an easy and extremely powerful criterion
for controllability and works well for finite-dimensional unitary gates since they are subgroups of
the compact group $SU(n)$. On the other hand the 
solutions of (\ref{eq:evolution}) no longer enjoy this property,
so we cannot apply the {\em rank criterion} directly.

Why and where is the compactness used to prove that the {\em rank criterion} is 
sufficient for controllability?
Given a generic Lie group $\mathcal{G}$, a set 
$\mathcal{E}=\{  \widetilde{G}_1,\cdots, \widetilde{G}_m\}$
of linearly independent generators of the corresponding Lie algebra,
and an element $K \in \mathcal{G}$,  one can write
\begin{align}
K &= e^{\widetilde{G}_1 t_1}e^{\widetilde{G}_2 t_2} \cdots e^{ \widetilde{G}_n t_n} \nonumber \\
& \textrm{with}  \:\: \widetilde{G}_j \in \mathcal{E}\:\: \textrm{and} \:\:
t_j \in \mathbbm{R}. \label{eq:rank}
\end{align}
In principle, in the expression above, there will be some exponentials involving
negative times $t_j$, while, if we want $K$ to be reachable by control, we need
all the times to be positive so that every exponential in the product corresponds
to an evolution described by Eq.  (\ref{eq:evolution}) and obtained by setting the 
control functions equal to certain constant values. As shown in \cite{JS,Dalessandro,Dalessandro2}, the compactness 
of the Lie group is a sufficient condition to switch the sign from negative to positive time: 
let $\widetilde{G}$ be one of the generators of $\mathcal{L}$ and consider 
a negative time $t <0$. 
If $\mathcal{G}=e^{\mathcal{L}}$ is compact,  then a sequence of positive times
$t_k>0$ always exists such that
\begin{align}
\lim_{k\rightarrow\infty} e^{\widetilde{G} t_k } = e^{ \widetilde{G} t} .
\end{align}
In other words, for a given $\epsilon>0$ and time $t<0$, 
we can always find a positive time $\tau>0$,
such that, {for a given matrix norm},
$\lVert e^{\widetilde{G}\tau}- e^{ \widetilde{G} t} \rVert <\epsilon $. \\
One should also notice that this condition is equivalent to saying that at a certain
time, the evolution operator recurs to the identity,
that is, 
recurrence is a necessary and sufficient condition to revert the sign
from negative to positive times. 
When one deals with non-compact groups, 
the possibility to switch from negative to positive times is in 
general lost. 
A simple and visually clear example in this sense is given by the squeezing operation 
in phase-space \cite{BR}: if we continuously apply the squeezing operation,
the state gets more and more squeezed and recurrence is never achieved. \\
However we will show in the following
that for coupled harmonic oscillators, 
even considering non-compact Lie groups, 
recurrence takes place with arbitrary precision if an additional condition
on the system's Hamiltonian is met.
Since, as we pointed out before, in the proof of the {\em  
rank criterion}, compactness is used only to revert the sign of time
in evolution operators, if one 
is able to achieve this goal by imposing other different physical and
mathematical constraints, then the {\em rank criterion} remains necessary
and sufficient to prove that the reachable set is dense in the Lie group
being considered.
\par
{\em Controllability of quadratic Hamiltonians} - 
Previously we showed that, given a system described
by Eq. (\ref{eq:Ham}), a linear control problem for the 
unitary operator $U$ is defined by the 
Schr\"odinger equation as in Eq. (\ref{eq:Uni}).
If we consider Hamiltonians bilinear in quadrature operators such that
\begin{align}
H_k = \frac{1}{2} \sum_{s,t} A_{st}^{(k)} R_s R_t  \qquad k = \{0,\dots,m\}, \label{eq:Ak}
\end{align}
the corresponding unitary operators $U$ are infinite-dimensional matrices; however 
we can consider the equivalent control problem,
with the form of Eq. (\ref{eq:evolution}), for the finite-dimensional evolution 
matrix $S$ as 
\begin{align}
\frac{dS}{dt} = -A(f,t)\Omega \: S\qquad S(0)=\mathbbm{1} \label{eq:finite}
\end{align}
where $A(f,t) = A^{(0)} + \sum_{k=1}^m f_k(t) A^{(k)}$.\\
The two evolution equations are equivalent and we will focus for the moment on 
the finite-dimensional representation in Eq. (\ref{eq:finite}). The symplectic group
$Sp(2n,\mathbbm{R})$ is a non-compact group and thus one cannot apply
the {\em rank criterion} to assess the controllability of the system. 
Our main result, contained in the following theorem, shows that if we can identify a set of 
linearly independent generators of the symplectic algebra 
$\mathcal{L}=sp(2n,\mathbbm{R})$,  such that the corresponding 
$\widetilde{A}^{(k)}$ are positive definite, 
one can achieve recurrence  with arbitrary precision, and thus the 
{\em rank criterion} will still be a sufficient condition for controllability.
\\
\par\noindent
{\bf Theorem}: If $A$ is  a positive (negative) definite matrix then, 
$$
\forall \: \epsilon >0\:\: \textrm{and} \:\: \forall \: T>0, 
\:\: \exists \: \tau>T \:\: \textrm{such that} \:\: \lVert e^{-A\Omega \tau} - \mathbbm{1}\rVert<\epsilon
$$ 
where we considered the Euclidean matrix norm 
$\lVert M \rVert = \sqrt{ \hbox{Tr}[M^\dag M]}$. \\
\noindent
{\bf Proof}:
 If $A$ is a positive definite matrix, because of Williamson theorem \cite{Williamson}, 
we can write $A=V D V^T$ where 
$
D = {\rm diag}\{ \nu_1,\nu_1,\dots, \nu_N,\nu_N\} $, $\nu_j \in \mathbbm{R}^+
$
and $V$ belongs to $Sp(2n,\mathbbm{R})$. 
This implies that  $V^T\Omega = \Omega V^{-1}$ and thus we obtain
$A\Omega = V D \Omega V^{-1}$.\\
The matrix $D\Omega$ is a normal matrix  diagonalized by a unitary matrix $U$,
such that $D\Omega=U D^\prime U^{\dag}$ with
$
D^\prime = {\rm diag} \{ +i\nu_1,-i\nu_1,\dots, +i\nu_N,-i\nu_N\}.
$ 
Then we have
\begin{align}
A\Omega &= V U D^{\prime} U^{\dag} V^{-1}  = W D^{\prime} W^{-1} 
\end{align}
that is $A\Omega$ has pure imaginary eigenvalues and is diagonalized by the
matrix $W=V U$. As a consequence, we can write the matrix $S(t)=e^{-A\Omega t}$ as
\begin{align}
S(t)= W  E(t) W^{-1}
\end{align}
with
$E(t) = {\rm diag} \{ e^{-i\nu_1}, e^{i\nu_1},\dots, e^{-i\nu_n},e^{i\nu_n}\}$,
which leads to 
\begin{align}
\lVert S(t) -\mathbbm{1} \rVert &= \lVert W(E(t) -\mathbbm{1}) W^{-1} \rVert \\
&\leq \lVert W \rVert\: \lVert W^{-1} \rVert \: \lVert E(t)-\mathbbm{1}\rVert.
\end{align}
The matrix $W$ is independent on time $t$, and thus 
$\lVert W \rVert\: \lVert W^{-1} \rVert  = K$ is constant.
Let us consider the remaining term
\begin{align}
\lVert E(t) -\mathbbm{1} \rVert &= \sqrt{ \hbox{Tr}[|E(t)-\mathbbm{1}|^2]} \\
&= \left (2 \sum_{k=1}^n |e^{-i\nu_k t} - 1|^2 \right)^{1/2}.
\end{align}
It is easy to check that $\lVert E(t) -\mathbbm{1} \rVert$ is a sum of trigonometric 
exponential functions and thus is a {\em quasi-periodic} function. Because of this property, 
a time $\tau>T$ exists, such that $\lVert E(\tau) -\mathbbm{1} \rVert$ is
close to zero with arbitrary precision \cite{bohr}. 
In particular, for a given $\epsilon>0$, we can choose a time $\tau$ such that 
$\lVert E(\tau) -\mathbbm{1} \rVert \leq \epsilon/K$  and
then obtain the thesis $\lVert S(\tau) -\mathbbm{1} \rVert \leq \epsilon$. $\blacksquare$ \\
It is worth also to notice that our result can be extended to semi-definite matrices $A$'s 
such that $\Omega A$ is diagonalisable, but it cannot be extended 
to generic semi-definite $A$'s. 
For instance, the Hamiltonian $H=p^2$ for a single degree of freedom would not 
recur.
\par
This theorem assures that if, by properly choosing the control functions $f_k(t)$
in the Hamiltonian given in Eq. (\ref{eq:Ham}), we can identify a set
of linearly independent  generators of the symplectic group, such that the corresponding 
matrices $\widetilde{A}^{(k)}$ are positive (negative) definite, then we can always revert the 
sign of negative times in the expression corresponding to  Eq. (\ref{eq:rank}). 
As a consequence  the {\em Lie algebra rank criterion}
 remains a necessary and sufficient 
condition to asses the controllability of the symplectic group, even if the group 
is not compact, and thus can be used to assess which Gaussian operations
can be realized given a certain control problem as indicated in Eq. (\ref{eq:finite}). \\
On a more fundamental level, our argument highlights a hitherto 
unnoticed connection between the normal mode decomposition of 
positive definite quadratic Hamiltonians -- formally an implication of the
Williamson theorem -- and the controllability of sets of coupled oscillators. 
This connection is bridged by the notion of dynamical recurrence
which, regardless of the infinite-dimensionality of the Hilbert space, is 
always guaranteed for positive definite quadratic Hamiltonians. 
The generality of our result makes it at first easy to overlook its potential impact: 
the applicability of quantum control to continuous-variable systems paves the way to 
vastly improving fidelities of current experiments as well as using control more 
efficiently. This fact can be clearly seen in the simple but quite surprising example 
we provide below, where our result is used to simplify the controllability properties 
of a large harmonic oscillator network.

\par
{\em Local controllability of a quadratic harmonic oscillator chain} - 
One of the most important requirements in quantum information and
quantum computation is to dynamically address and control individual
interacting systems. From an experimental point of view, it is also desirable 
to have complete control on a large network by acting only on a small
part of it. Indirect control has been already proved for qubit systems \cite{Daniel}, 
while, regarding networks of harmonic oscillators, it has been recently shown, for instance, 
that by probing only one site of the network, one can reconstruct the full
quantum state of the system \cite{tom}.
Here we use our theorem to prove the controllability 
of a chain of harmonic oscillators, where only one or few sites of the 
chain are accessible.
This kind of example is relevant if we think, for example, of
an array of interacting trapped ions, where in principle one can implement 
Gaussian operations by addressing single ions and manipulating their trapping 
frequencies 
\cite{ions2,alessioIons}.\\
Let us start by defining 
the bosonic mode operators $a_j = (q_j+i p_j)/\sqrt{2}$ and
$a_j^\dag = (q_j-i p_j)/\sqrt{2}$, satisfying
the commutation relation $[a_i, a_j^{\dag}]=\delta_{ij}$. 
Let us consider an $n$-mode bosonic chain, described by a Hamiltonian
as in Eq. (\ref{eq:Ham}). In particular the always-on Hamiltonian reads
\begin{align}
H_0 &= \omega \sum_{j=1}^n  \left( a_j^{\dag} a_j  + \frac12 \right) 
+ g_1 \sum_{j=1}^{n-1} (a_j a_{j+1}^\dag + h.c.) + 
\nonumber \\
&\:\:\: + g_2 \sum_{j=1}^{n-1} (a_j a_{j+1} + h.c.).   \label{eq:H0}
\end{align}
where, for the sake of simplicity, we consider all the oscillators having
with the same frequency $\omega$. If we consider $g_1=g_2$, this 
corresponds to the $q_j q_{j+1}$ coupling, which is the most common
between the harmonic oscillators' interactions. 
From now on we will consider
the renormalized coupling constants $\widetilde{g}_j=g_j/\omega$ and 
assume that they are both positive; a sufficient condition for 
the positivity of $H_0$ (for every number of bosons in the chain $n$) is
$\widetilde{g}_1+\widetilde{g}_2 < 1/2$. 
 We consider as the controlling Hamiltonians, 
a local {\em phase-rotation} and a local {\em squeezing} 
term on the first mode of the chain only, {\em i.e.}:
\begin{align}
H_1= \omega_1 \left( a_1^\dag a_1  \right) \:\:\: \textrm{and} \:\:\;
H_2 = \chi (a_1^2 + a_1^{\dag 2} )\: .
\label{eq:Hk}
\end{align}
We prove that, by denoting with $\mathcal{L}$ the symplectic
algebra,
\begin{align}
\mathcal{L}= \langle i H_0, iH_1,iH_2 \rangle_{[\cdot,\cdot]},
\end{align}
that is, by computing all possible commutators of these operators, of any order,
and their linear combinations, we can obtain all the elements of $\mathcal{L}$
(details of the proof can be found in the Supplemental Material).
Since the Lie algebra is a vector space, any set of linearly independent
linear combinations of the above operators satisfies the {\em rank criterion}. 
We have then to show that, by properly setting the control functions $f_k(t)$,
we can identify one of these sets, such that the corresponding matrices $\widetilde{A}^{(k)}$
are positive definite. There are in principle infinite choices, in particular it is 
easy to check that the following set fulfils all the conditions above
\begin{align}
\widetilde{H}_0 &= H_0 \\
\widetilde{H}_1 &= H_0 + \alpha H_1 \qquad &\alpha\omega_1> 0 \\
\widetilde{H}_2 &= H_0 + \beta H_1 + \delta H_2 \qquad &0<\delta \chi<\beta \omega_1 \: .
\end{align}
\par
In practice, this example is directly relevant to arrays of trapped ions
and chains of nano-mechanical oscillators. For instance, in the case of
the transverse ionic modes, local controls analogous to (\ref{eq:Hk}) could be
obtained by manipulating the local trapping frequencies, as detailed in
\cite{alessioIons} and realised in \cite{ions2} (analogous forms of control 
have been envisaged for opto-mechanical setups as well \cite{mari}).
Note also that even restricted
local control might suffice for certain manipulations, depending on the
desired tasks. For instance, as a side product of the proof reported in
the Supplemental Material, one can show that, if the local control is restricted to phase
rotations generated by $H_1$, the whole symplectic algebra is not
achievable, but all the passive operations, comprising beam-splitters and
local phase rotations can be realized. This would allow one to implement,
for example, cooling protocols based on swapping excitations between 
sites of the array. 
\par
{\em Conclusions} - 
While the general control theory of infinite-dimensional systems remains hard,
we have found a surprisingly simple solution for the case of quadratic interactions
of coupled harmonic oscillators. To demonstrate the applicability of our result, 
we have also discussed its application to indirect control (and, potentially, cooling) 
of chains of oscillators.\\
It is worth mentioning that proving controllability and finding an actual control pulse
are completely distinct tasks. For instance, using the quantum recurrence theorem
on the theory level is useful, but for a control pulse one could not rely on it,
as the recurrence and hence the resulting pulses would take far too much time.
This is well understood and typically overcome by using numerical routines to optimize
the pulses. 
In our case, a similar point arises regarding the requirement of positive (negative)
definiteness of the Hamiltonian, which was used as a sufficient element of the proof.
In an actual pulse sequence, it could be beneficial to use non-positive or negative definite 
Hamiltonians in order to achieve faster control. \\
Last, but not least, let us remark that the dynamics of classical systems governed by quadratic Hamiltonians  
is also described by the symplectic group of canonical transformations: our finding hence applies, as it stands, to the controllability of classical, as well as quantum harmonic oscillators.
\par
{\em Acknowledgements} - 
MSK acknowledges support from UK EPSRC and MGG acknowledges 
a fellowship support from UK EPSRC (grant EP/I026436/1) . 
Part of this work was carried out while DB held the UK EPSRC grant EP/F043678/1 
at Imperial College London.

\clearpage
\section{Supplemental material}
\subsection{Proof of rank criterion for harmonic oscillator chain}\label{app:rank}
Here we consider the control problem defined by the Hamiltonian
\begin{equation}
H(t) = H_0 + \sum_{k=1}^m f_k(t) H_k, \label{eq:Ham}
\end{equation}
where the always-on Hamiltonian reads
\begin{align}
H_0 &= \omega \sum_{j=1}^n  \left( a_j^{\dag} a_j  + \frac12 \right) 
+ g_1 \sum_{j=1}^{n-1} (a_j a_{j+1}^\dag + h.c.) + 
\nonumber \\
&\:\:\: + g_2 \sum_{j=1}^{n-1} (a_j a_{j+1} + h.c.),   \label{eq:H0}
\end{align}
the local controlling Hamiltonians are
\begin{align}
H_1= \omega_1 \left( a_1^\dag a_1  \right) \:\:\: \textrm{and} \:\:\;
H_2 = \chi (a_1^2 + a_1^{\dag 2} )\: ,
\label{eq:Hk}
\end{align}
and $a_j$ and $a_j^\dagger$ are $n$ pairs of bosonic operators satisfying 
the commutation relation $[a_j, a_k] = \delta_{j,k}$.\\
In the following we prove that the {\em Lie algebra rank criterion} for this system 
is fulfilled, that is,  by denoting with $\mathcal{L}$ the symplectic
algebra, then
\begin{align}
\mathcal{L}= \langle i H_0, iH_1,iH_2 \rangle_{[\cdot,\cdot]}.
\end{align}
We will work with the infinite-dimensional representation of the 
real symplectic group, where the basis of the corresponding Lie algebra
reads
\begin{align}
\mathcal{L} = &\{ i (a_j a_k^\dag + a_j^\dag a_k), a_j a_k^\dag - a_j^\dag a_k,,\nonumber \\
& i (a_j^\dag a_k^\dag + a_j a_k), a_j^\dag a_k^\dag - a_j a_k \}
\nonumber \\
&\textrm{with} \:\:  j,k =1,\dots,n  \label{eq:algebra} \: .
\end{align}
Proving the {\em Lie algebra rank criterion} corresponds to showing that by computing
all possible commutators of the control operators $H_k$, of any order and 
their linear combinations, we can generate all the elements of the basis
of the algebra listed above. For the sake of simplicity we will prove it
by considering the case $g_1=g_2$ 
in the Hamiltonian $H_0$  (\ref{eq:H0}); 
this corresponds to considering a $q_j q_{j+1}$ coupling
between the sites of the chain, while the most general case 
(which comprises also the {\em rotating-wave approximation}
case where $g_2=0$ ) can be proved following the same line of reasoning.
We start by showing that we can generate
all the elements corresponding to the two first sites of the chain:
\begin{align}
S_a^{(1)}&=\frac12 [ i H_2, i H_1] = a_1^{\dag 2} - a_1^2 \\
O_1 &= [ i H_0, i H_1] = a_1^\dag a_2^\dag + a_1^\dag a_2 - a_1 a_2^\dag - a_1 a_2 \\
O_2 &= [ i H_1, O_1] = i(  a_1^\dag a_2^\dag + a_1^\dag a_2 + a_1 a_2^\dag + a_1 a_2 ) \\
R^{(12)}_a &= \frac12 \left [  \left( [ O_1, i H_0]  + 2 O_2 \right), i H_1 \right] = a_1^\dag a_2 - a_1 a_2^\dag \\
 T_a^{(12)} &= O_1 - R^{(12)}_a = a_1^\dag a_2^\dag - a_1 a_2 \\
 S_a^{(2)} &= [ T_a^{(12)}, R_a^{(12)} ] +S_1^{(1)}= a_2^{\dag 2} - a_2^2 \\
 T_b^{(12)} &= [ O_1 - R_a^{(12)} , i H_1] = i (a_1^\dag a_2^\dag + a_1 a_2 )\\
 O_3 &= [ i H_1 , O_1-2 R_a^{(12)} ] \nonumber \\
 &= i  ( a_1^\dag a_2^\dag - a_1^\dag a_2 - a_1 a_2^\dag + a_1 a_2 ) \\
 R_b^{(12)} &= \frac12 ( O_2 - O_3) = i (a_1^\dag a_2 + a_1 a_2^\dag) \\
 P^{(2)} &= \frac12 \left( 2i H_1 + [ R_b^{(12)}, R_a^{(12)}]   \right) = i a_2^\dag a_2 \\
 S_b^{(2)} &= [ P^{(2)}, S_a^{(2)} ] = i (a_2^{\dag 2} + a_2^2 ) \: .
\end{align}
The operators $P^{(2)}$, $S_a^{(1)}$, $S_a^{(2)}$, $S_b^{(2)}$, $T_a^{(12)}$, 
$T_b^{(12)}$, $R_a^{(12)}$, $R_b^{(12)}$, together
with the local control operators $i H_k$, complete all the operators corresponding
to the first two sites. Because of the symmetry of $H_0$, we can proceed 
in the same way for all the operators belonging to the neighbouring sites.
Then, to obtain the two-mode {\em long-distance} operators,
one can easily show that
\begin{align}
[ i (a_k^\dag a_l + a_k a_l^{\dag}), a_j a_k^\dag - a_j^\dag a_k ] = i (a_j^\dag a_l + a_j a_l^\dag) 
\end{align}
and analogue commutators for the remaining terms. The proof is hence complete. $\blacksquare$

\end{document}